\documentclass[aps,showpacs,preprintnumbers,amsmath,amssymb,superscriptaddress,floatfix,nofootinbib]{revtex4}
\usepackage{graphicx}
\usepackage{epsfig}
\usepackage{epstopdf}
\usepackage{bm}
\usepackage{amsmath}
\allowdisplaybreaks[1]
\usepackage{amsfonts}
\usepackage{amssymb}
\usepackage{multirow}
\usepackage{makecell}
\usepackage{slashed}
\usepackage[dvipsnames]{xcolor}
\usepackage{ulem}
\usepackage{longtable}
\usepackage{array}
\usepackage{hyperref}
\newcommand{\Lag}{\mathcal{L}}
\newcommand{\mP}{\mathcal{P}}
\newcommand{\mM}{\mathcal{M}}

\usepackage{subfigure}

\begin{document}

\title{Strange molecular partners of $P_c$ states in $\gamma p\to\phi p$ reaction}

\author{Shu-Ming Wu}\email{wushuming@mail.itp.ac.cn}
\affiliation{CAS Key Laboratory of Theoretical Physics, Institute
	of Theoretical Physics,  Chinese Academy of Sciences, Beijing 100190, China}
\affiliation{School of Physical Sciences, University of Chinese Academy of Sciences, Beijing 100049, China}

\author{Fei Wang}\email{wangfei@mails.ucas.ac.cn}
\affiliation{CAS Key Laboratory of Theoretical Physics, Institute
	of Theoretical Physics,  Chinese Academy of Sciences, Beijing 100190, China}
\affiliation{School of Physical Sciences, University of Chinese Academy of Sciences, Beijing 100049, China}

\author{Bing-Song Zou}\email{zoubs@mail.itp.ac.cn}
\affiliation{CAS Key Laboratory of Theoretical Physics, Institute
	of Theoretical Physics,  Chinese Academy of Sciences, Beijing 100190, China}
\affiliation{School of Physical Sciences, University of Chinese Academy of Sciences, Beijing 100049, China}
\affiliation{School of Physics, Peking University, Beijing 100871, China}

\begin{abstract}
Based on the high statistical data of the CLAS Collaboration on $\gamma p \to \phi p$ reaction in the center-of-mass energy range of 2.2 GeV to 2.8 GeV, we investigate the possible existence of strange molecular partners of $P_c$ states, i.e., $N^*(2080)$ and $N^*(2270)$ as $K^*\Sigma$ and $K^*\Sigma^*$ molecular states.
In addition to the t-channel Pomeron exchange, t-channel meson exchange including pseudo-scalar meson $(\pi,\eta)$, scalar meson $(\sigma, a_0(980), f_0(980))$, axial-vector meson $f_1(1285)$, tensor meson $f_2(1270)$, as well as s- and u-channel proton exchange,  including s-channel $N^*(2080)$ and $N^*(2270)$ states can fit the data very well. 
The fitted coupling constants of these $N^*$ molecular states to $p\phi$ and $\gamma p$ are consistent with the results directly calculated from the relevant hadronic triangle diagrams of the molecular picture.

\end{abstract}

\maketitle

\section{Introduction}

The observation of three narrow $P_c$ states decaying to $J/\psi p$ by LHCb experiment~\cite{LHCb:2015yax,LHCb:2019kea} has triggered a great interests and a lot of efforts to understand their nature~\cite{Chen:2016qju,Guo:2017jvc,Liu:2019zoy}.
In fact, the observed three narrow $P_c$ states, {\sl i.e.}, $P_c(4312)$, $P_c(4440)$ and $P_c(4457)$, are in consistent with earlier predictions~\cite{Wu:2010jy,Wu:2010vk,Wang:2011rga,Wu:2019adv} of one narrow $\bar D\Sigma_c$ bound state with spin-parity $J^P={\frac12}^-$ and two nearly degenerate narrow $\bar D^*\Sigma_c$ bound states with $J^P={\frac12}^- \& {\frac32}^-$. 
From heavy quark spin symmetry, one $\bar D\Sigma^*_c$ bound state with $J^P={\frac32}^-$ and three $\bar D^*\Sigma^*_c$ bound states with $J^P={\frac12}^-, {\frac32}^- \& {\frac52}^-$ are also expected to exist~\cite{Xiao:2013yca,Liu:2019tjn,Du:2019pij}. 
Evidence of a $P_c(4380)$ was claimed by Refs.\cite{LHCb:2015yax,Du:2019pij}, with its mass in consistent with the expectation of the $\bar D\Sigma^*_c$ bound state. 
Since these hidden-charm $P_c$ states can be successfully and naturally ascribed by the hadronic molecular picture, their strange partners are also expected to exist~\cite{He:2017aps,Zou:2018uji,Lin:2018kcc,Ben:2023uev}.  
As the $P_c$ states were observed through their $J/\psi p$ decay mode, for those of their strange partners above $\phi p$ threshold, they are expected to show up in the corresponding $\phi p$ decay mode. 
This prompts us to look for the $K^*\Sigma$ and $K^*\Sigma^*$ bound states in  $\gamma p\to\phi p$ reaction.

In 2014, the CLAS Collaboration at Jefferson Lab has reported high statistics measurements of differential cross sections for the reaction $\gamma p \to \phi p$~\cite{CLAS:2013jlg,Dey:2014tfa}.
The experimental results show that there may be some structures near the center-of-mass energies $W=2.1\  \text{GeV}$ and $2.3\ \text{GeV}$.
And there is a bump structure at the forward angle around $W=2.1\ \text{GeV}$.

Before the results of this experiment were released, there have been many studies on this process~\cite{Zhao:1999af,Titov:1999eu,Oh:2001bq,Zhao:2001ue,Titov:2007fc,Ozaki:2009mj,Kiswandhi:2010ub,Ryu:2012tw}.
Due to the insufficient statistics of the previous experimental data, the previous models need to be improved accordingly when describing the latest experimental results.

Then based on the latest CLAS data, B.G. Yu et al.~\cite{Yu:2016zut} used a Reggeized parameterized meson $(\pi,\sigma,f_2)$ exchange and Pomeron exchange to explain this forward angle behavior.

S.H. Kim and his collaborators~\cite{Kim:2019kef,Kim:2020wrd} considered t-channel Pomeron exchanges, t-channel meson exchanges including ($\pi,\eta,a_0, f_0,f_1(1285)$), s-channel and u-channel proton exchange and s-channel nucleon resonances exchange including ($N^*(2000,5/2^+),N^*(2300,1/2^+)$) in PDG~\cite{Workman:2022ynf}.
In later work~\cite{Kim:2021adl}, they also considered the $\phi N \to \phi N$ final state interaction including the gluon-exchange interaction, the direct $\phi N$ coupling term and coupled-channel effects arising from the one-meson-exchange mechanisms in $\phi N \to K \Lambda,K \Sigma, \pi N, \rho N$ processes, and found that the effect of final state interaction of $\gamma p \to \phi p$ is very small.

Since in $W=2.1\sim 2.3\ \text{GeV}$ energy region for the $\gamma p \to \phi p$ reaction, there may be two hadronic molecular states: S-wave $K^*\Sigma$ molecule $N^*(2080)$ and S-wave $K^*\Sigma^*$ molecule $N^*(2270)$.
Then, we examine whether the data can be fitted by using $N^*(2080)$ and $N^*(2270)$ instead of previous $N^*(2000,5/2^+)$ and $N^*(2300,1/2^+)$ for the s-channel $N^*$ exchange together with the above mentioned background terms.
%

The article is organized as follows.
In Sec.~\ref{sec:formalism}, we present the theoretical framework of our calculation.
Our results are shown in  Sec.~\ref{sec:result} as well as some discussions and a brief summary. Appendix~\ref{AppendixA} is presented at last.

\section{Theoretical framework} \label{sec:formalism}

For the reaction of $\phi$ photoproduction: $\gamma(k_1) + p(p_1) \to \phi(k_2) + p(p_2)$ in the center-of-mass system, the four-momenta of these particles can be defined as:
\begin{eqnarray}
	k_1 &=& (k,\vec{k}) ,\\
	p_1 &=& (E_{p}(k),-\vec{k}) ,\\
	k_2 &=& (E_{\phi}(k'), \vec{k}') ,\\
    p_2 &=& (E_{p}(k'),-\vec{k}') ,\\
    P &=& k_1 + p_1 = k_2 +p_2 = (W,\vec{0}),
\end{eqnarray}
where $k (k')$ is the magnitude of three-momenta $\vec{k} (\vec{k}')$, $E_a(k) = \sqrt{M_a^2+k^2}$ is the energy of a particle with mass $M_a$, and $W$ is the invariant mass of the system.

Then the differential cross section can be expressed as follows:
\begin{equation}
 \frac{d\sigma}{d\Omega} = \frac{1}{64 \pi^2 W^2} \frac{k'}{k}\frac{1}{4} \sum _{all \ spins} 
 |M_{\lambda_1,s_2,\lambda_2,s_2}(k_1,p_1,k_2,p_2)|^{2},
	\label{eq:dsigma}
\end{equation}
where the invariant amplitude $M$ can be written as:
\begin{equation}
	M_{\lambda_1,s_2,\lambda_2,s_2}(k_1,p_1,k_2,p_2) = \epsilon_{\nu}^{*}(k_2,\lambda_2) 
 \bar{u}(p_2,s_2) \mM^{\mu\nu}(k_1,p_1,k_2,p_2) u(p_1,s_1) \epsilon_{\mu}(k_1,\lambda_1),
\end{equation}
where $\epsilon_\mu(k_1,\lambda_1)$ and $\epsilon_\nu^*(k_2,\lambda_2)$ are the polarization vector of photon $\gamma$ and meson $\phi$, respectively,
$u(p_1,s_1)$ and $\bar{u}(p_2,s_2)$ are the spinor of the incoming and outgoing baryon, respectively, 
with the normalization $\bar{u}(p,s) u(p,s') = 2M_p\delta_{s,s'}$.

\begin{figure}[htbp]
\centering
\subfigure[t-channel pomeron exchange]{
	\includegraphics[width=0.4\linewidth]{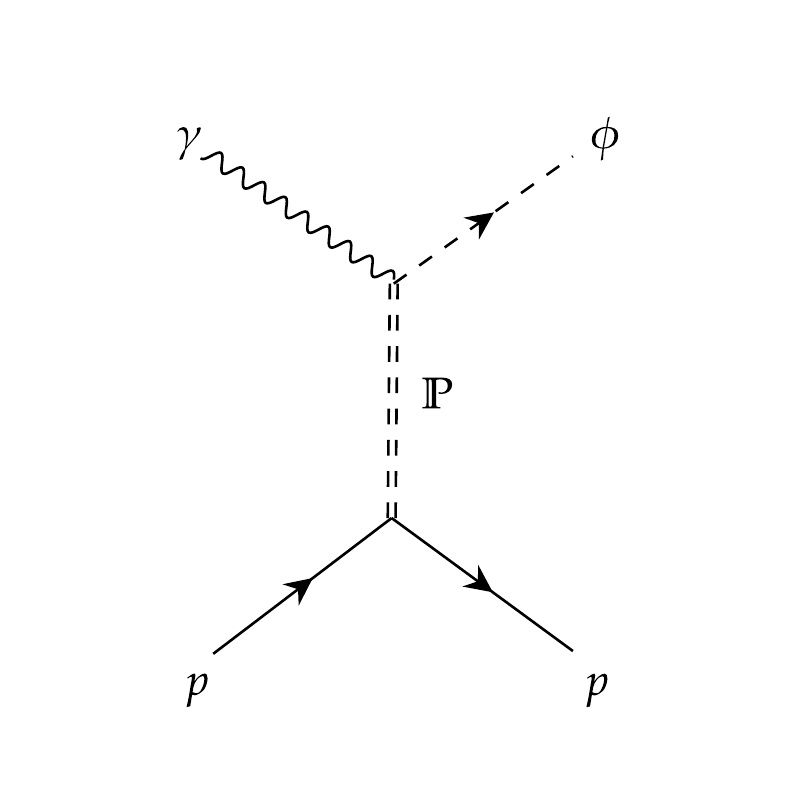}}
\subfigure[t-channel mesons exchange]{
	\includegraphics[width=0.4\linewidth]{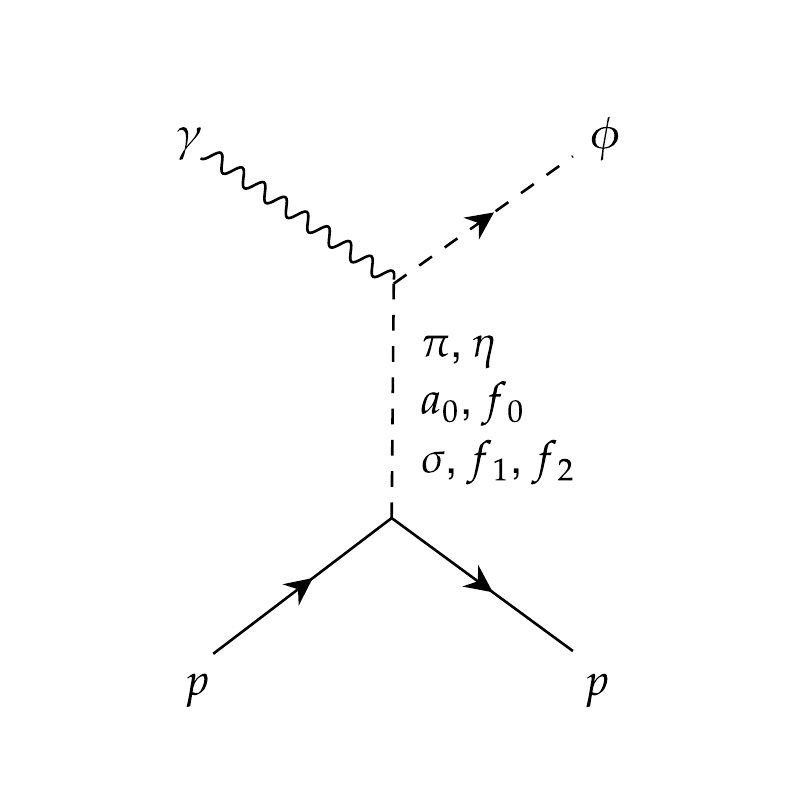}}
\\
\subfigure[u-channel proton exchange]{
	\includegraphics[width=0.4\linewidth]{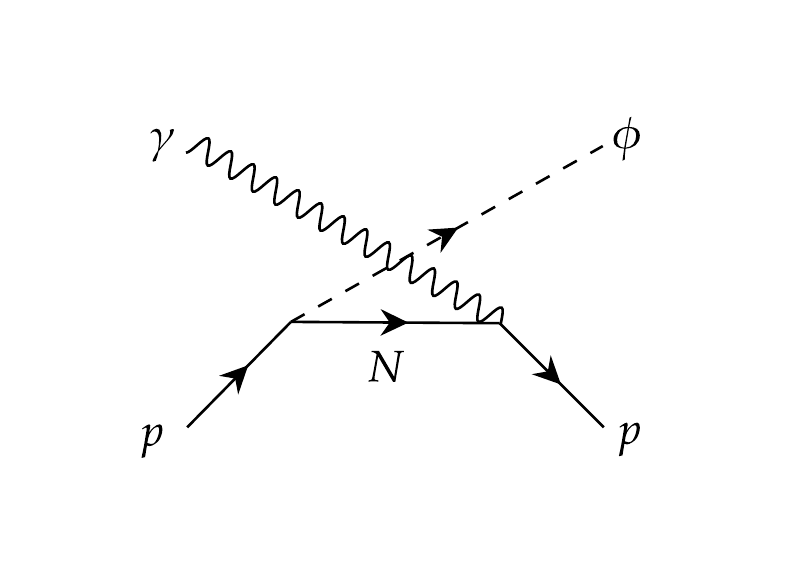}}
\subfigure[s-channel nucleon exchange]{
	\includegraphics[width=0.4\linewidth]{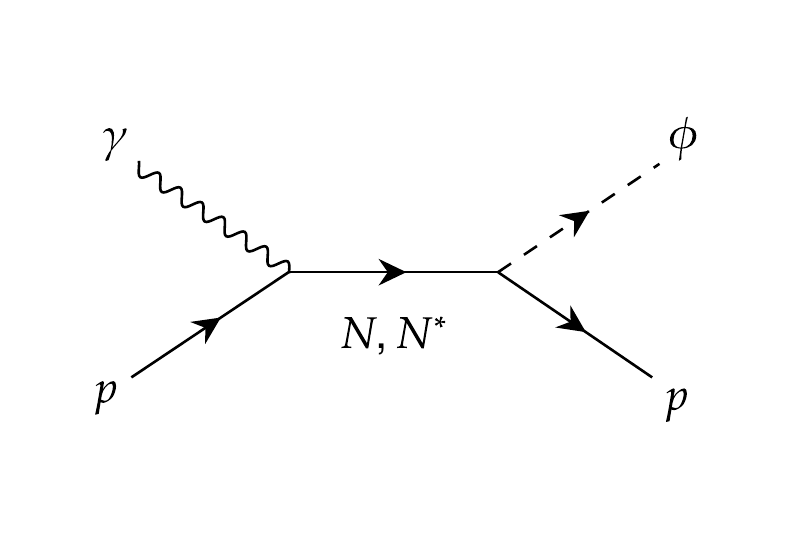}}
\caption{Relevant Feynman diagrams for $\gamma p \to \phi p$.}
\label{fig:diagrams}
\end{figure}

As shown in Fig.~\ref{fig:diagrams}, the full amplitude in our model consists of t-channel Pomeron exchange, t-channel meson exchange including pseudo-scalar meson $(\pi,\eta)$, scalar meson $(\sigma, a_0(980), f_0(980))$, axial-vector meson $(f_1(1285))$, tensor meson $(f_2(1270))$, s- and u-channel proton exchange, and s-channel $N^*$ molecule exchange. 
Then, $\mM^{\mu\nu}$ can be written as:
\begin{eqnarray}
    \mM^{\mu\nu} &=& \mM^{\mu\nu}_{t-ch,\mathbb{P}} + \mM_{t-ch,M}^{\mu\nu} + \mM_{N}^{\mu\nu} + \mM_{s-ch,N^*}^{\mu\nu}.
\end{eqnarray}

In the following we present these amplitudes in detail.
\subsection{Pomeron exchange}

\begin{figure}[htbp]
	\centering
	\includegraphics[width=0.6\linewidth]{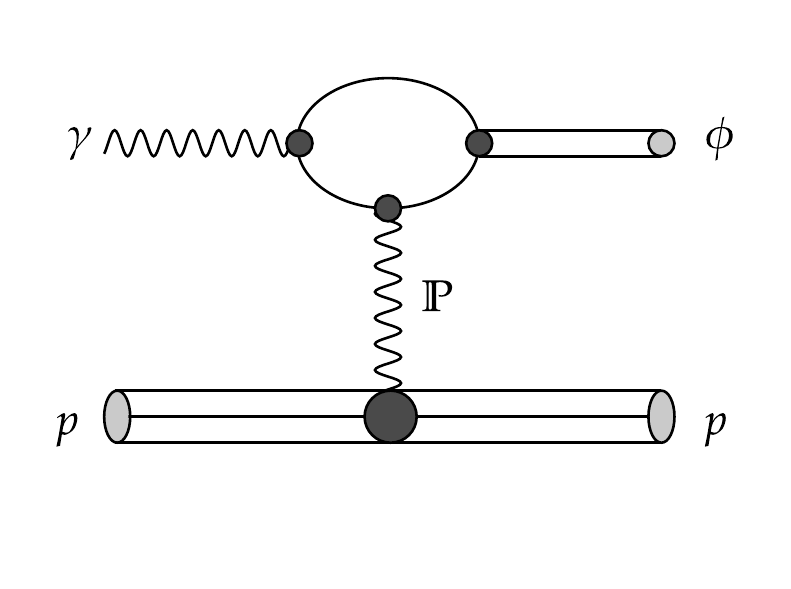}
	\caption{\label{fig:pomeron_mechanism}The Pomeron exchange mechanism of $\gamma p \to \phi p$ in quark level.}
\end{figure}

The Pomeron exchange based on the Regge phenomenology is one of the most successful approaches to high energy elastic scattering. 
In the study of Donnachie and Landshoff~\cite{Donnachie:1984xq}, it can be approximated by considering it as a particle with $I(J^C)=0(1^+)$, which mainly couples to the quarks in hadrons. 
This is shown schematically in Fig.~\ref{fig:pomeron_mechanism}.
And the Pomeron couples to quarks with $\gamma_\mu$ type, similar to a photon.

After considering some approximations~\cite{ Laget:1994ba, Pichowsky:1996tn, Zhao:1999af}, the Pomeron exchange amplitude can be written as:
\begin{eqnarray}\label{eq:MP}
    \mathcal{M}^{\mu\nu}_{t-ch,\mathbb{P}}(k_1,p_1,k_2,p_2) = G_\mathbb{P}(s,t) \mathcal{T}^{\mu\nu}_\mathbb{P}(k_1,p_1,k_2,p_2) ,
\end{eqnarray}
with $s=P^2=W^2$, $t=(p_1-p_2)^2=(k_1-k_2)^2$,
\begin{eqnarray}
    \mathcal{T}^{\mu\nu}_{\mathbb{P}}(k_1,p_1,k_2,p_2) = i 12\frac{e M_\phi^2}{f_\phi} \beta_{s}F_\phi(t)\beta_{u/d} F_1(t)
    \left( \rlap{\,/}{k_1} g^{\mu\nu} - k_1^\nu \gamma^\mu \right)  \, ,
    \label{eq:pom-a}
\end{eqnarray}
where the decay constant $f_\phi = 13.48$ can be calculated from the decay width of $ \phi \to e^+ e^-$ in VMD model~\cite{Wu:2019adv}.
And $\beta_{s}$ ($\beta_{u/d}$) is the coupling constant of the Pomeron with the quarks $s$ ($u$ or $d$) in the vector meson $\phi$ (proton $p$).
The $F_\phi(t)$ and $F_1(t)$ are the form factor for the Pomeron-vector meson vertex and isoscalar electromagnetic form factor of the nucleon, respectively, which can be expressed as follows:
\begin{eqnarray}
    F_\phi(t) &=& \frac{1}{M_\phi^2-t} \left( \frac{2\mu_0^2}{2\mu_0^2 + M_\phi^2 - t} \right),\\
    F_1(t) &=& \frac{4M_p^2 - 2.8 t}{(4M_p^2 - t)(1-t/0.71\text{GeV}^2)^2},
    \label{eq:f1v}
\end{eqnarray}
where $\mu_0$ is a cutoff of the form factor related to the Pomeron-vector meson vertex.

The $G_\mathbb{P}$ in Eq.~(\ref{eq:MP}) is the Regge propagator of the Pomeron, and it is written as follows:
\begin{eqnarray}
    G_\mathbb{P} = \left(\frac{s}{s_0}\right)^{\alpha_P(t)-1}\exp\left\{ - \frac{i\pi}{2} \left[ \alpha_P(t)-1 \right] \right\} ,
    \label{eq:regge-g}
\end{eqnarray}
with Regge trajectory $ \alpha_P (t) = \alpha_0 + \alpha'_P t$.

The parameters $\mu_0^2, \beta_{u/d}, \beta_{s}, s_0, \alpha_0, \alpha'_P$ can be determined by fitting the total cross section of $\rho_0$, $\omega$ and $\phi$ photoproduction at high energies~\cite{Oh:2000zi}. 
Here we use the same values as in Refs.~\cite{Wu:2019adv,Kim:2021adl}:
\begin{gather}
    \mu_0^2 = 1.1\:\text{GeV}^2,\quad \beta_{u/d} = 2.07\:\text{GeV}^{-1},\quad \beta_{s} = 1.386\:\text{GeV}^{-1},\nonumber\\
    \alpha_0 = 1.08,\quad \alpha'_P = 1/s_0 = 0.25\:\text{GeV}^{-2}.\nonumber
\end{gather}

\subsection{Meson exchange}

In the low-energy region, we also consider t-channel common meson exchanges, including pseudo-scalar meson $\varphi=\{\pi,\eta\}$, scalar meson $S=\{\sigma, a_0(980), f_0(980)\}$, axial-vector meson $f_1(1285)$ and tensor meson $f_2(1270)$.
Corresponding effective Lagrangians are given as in Refs.~\cite{Kim:2019kef,Kim:2021adl}:
\begin{align}
    \Lag_{\gamma \varphi \phi} &= \frac{e g_{\gamma \varphi \phi}}{M_\phi} \epsilon^{\mu\nu\alpha\beta} \partial_\mu A_{\nu} \partial_\alpha V_{\beta} ,\label{eq:ml_gammaphiphi} \\[3pt]
    \Lag_{\gamma S \phi} &= \frac{e g_{\gamma S \phi}}{M_\phi} F^{\mu\nu} \phi_{\mu\nu}S ,\label{eq:ml_gammaSphi} \\[3pt]
    \Lag_{\gamma f_1 \phi} &=  
    g_{\gamma f_1 \phi} \epsilon^{\mu\nu\alpha\beta} \partial_\mu A_\nu \phi_\alpha f_{1\beta} ,\label{eq:ml_gammaAphi} \\[3pt]
    \Lag_{\gamma f_2 \phi} &=  \frac{g_{\gamma f_2 \phi}}{m_0} F_{\mu\alpha} g^{\alpha\beta} \phi_{\beta\nu} f_2^{\mu\nu}, \\[3pt]
    \Lag_{\varphi N N} &= -i g_{\varphi  N N}\bar{N}\gamma_5 N \varphi ,\label{eq:ml_phiNN} \\[3pt]
    \Lag_{S N N} &= g_{S N N}\bar{N} N S ,\label{eq:ml_SNN} \\[3pt]
    \Lag_{f_1 N N} &= -g_{f_1 N N}\bar{N} \gamma_\nu f_1^\mu \gamma_5 N ,\label{eq:ml_ANN} \\[3pt]
    \Lag_{f_2 N N} &= \frac{2g_{f_2 N N}}{M_N} [\bar{N}\gamma_\mu,\partial_\nu]N f_2^{\mu\nu} ,\label{eq:ml_TNN}
\end{align}
where $\varphi=\{\pi,\eta\}$, $S=\{\sigma,a_0,f_0\}$, $F^{\mu\nu}=\partial^{\mu}A^{\nu}-\partial^{\nu}A^{\mu}$, $\phi_{\mu\nu}=\partial_{\mu}\phi_{\nu}-\partial_{\nu}\phi_{\mu}$.

The amplitude can be calculated from the above effective Lagrangians:
\begin{align}
	\mM_{t-ch,\varphi}^{\mu\nu} =&\  i \frac{e g_\varphi} {M_\phi} \epsilon^{\mu\nu\alpha\beta} k_{1\alpha} k_{2\beta} \gamma_5 \frac{F_{\varphi}(t)}{t-M_\varphi^2} ,\\
	\mM_{t-ch,S}^{\mu\nu}  =&\  2 \frac{e g_S}{M_\phi} (k_1 \cdot k_2 g^{\mu\nu} - k_1^\nu k_2^\mu) \frac{F_{S}(t)}{t-M_S^2} ,\\
	\mM^{\mu\nu}_{t-ch,f_1}  =&\  i 
	g_{f_1} \epsilon^{\mu\nu\alpha\beta}\left[ -g_{\alpha\lambda}+\frac{q_{t\alpha} q_{t\lambda}}{M_{f_1}^2} \right] \nonumber\\
	&\times \gamma^\lambda \gamma_5 k_{1 \beta} \frac{F_{f_1}(t)}{t-M_{f_1}^2},\\
	M^{\mu\nu}_{t-ch,f_2}  =&\ \Gamma_{\gamma f_2 \phi}^{\mu \nu \beta \rho} G^2_{\beta\rho;\lambda\sigma}(q_t) \Gamma_{f_2 N N}^{\lambda \sigma} \frac{F_{f_2}(t)}{t-M_{f_2}^2},
\end{align}
with $g_a = g_{\gamma a\phi} g_{aNN}\: (a=\{\varphi,S,f_1\})$,
\begin{align}
    \Gamma_{\gamma f_2 \phi}^{\mu \nu \beta \rho}(k_1, k_2) =&\ \frac{g_{\gamma f_2 \phi}}{m_0}\left( k_1^\beta k_2^\rho g^{\mu\nu} + k_1 \cdot k_2 g^{\nu\beta} g^{\mu\rho}\right. \nonumber\\
    & \left.- k_1^{\nu} k_2^\rho g^{\mu\beta} - k_1^\rho k_2^{\mu} g^{\nu\beta} \right) ,\\
    G^2_{\beta\rho;\lambda\sigma}(q) =&\  \frac{1}{2}\left(\tilde{g}_{\beta\lambda} \tilde{g}_{\rho\sigma} + \tilde{g}_{\beta\sigma} \tilde{g}_{\lambda\rho}\right) - \frac{1}{3} \tilde{g}_{\beta\rho}\tilde{g}_{\lambda\sigma},\\
    \Gamma_{f_2 N N}^{\lambda \sigma}\left(p_1, p_2\right) =&\ \frac{2 g_{f_2 N N}}{M_N}(p_1+p_2)^\lambda \gamma^\sigma ,
\end{align}
with $m_0 = 1.0\:\text{GeV}$, $\tilde{g}^{\beta \rho}=-g^{\beta \rho}+q^\beta q^\rho/m_{f_2}^2 $.
$F_{\varphi}(t)$, $F_{S}(t)$, $F_{f_1}(t)$  and $F_{f_2}(t)$ are off-shell form factors taken as :
\begin{eqnarray}
	F_{\varphi,S}(t) &=& e^{i\beta_{\varphi,S}}\frac{\Lambda_{\varphi,S}^4}{\Lambda_{\varphi,S}^4+(t-M_{\varphi,S}^2)^2} ,\\
	F_{f_1,f_2}(t) &=& e^{i\beta_{f_1,f_2}} \left( \frac{\Lambda_{f_1,f_2}^4}{\Lambda_{f_1,f_2}^4+(t-M_{f_1,f_2}^2)^2} \right)^2.
\end{eqnarray}	

In order to better describe the experiment, we use Regge theory to deal with the $\sigma$ exchange, the $f_1$ exchange and the $f_2$ exchange, which can be referred to Refs.~\cite{Yu:2016zut,Kim:2019kef}. Then we replace the Feynman propagator $1/(t-m_{\sigma}^2)$, $1/(t-m_{f_1}^2)$  and $1/(t-m_{f_2}^2)$ with Regge propagator:
\begin{align}
    \frac{1}{t-m_{\sigma}^2} \to \mathcal{P}_{\sigma}(s,t) =&\  \frac{\pi \alpha_{\sigma}'\times D_{\sigma}(t)}{\Gamma[\alpha_{\sigma}(t)+1]\sin{[\pi \alpha_{\sigma}(t)}]}\times \left(\frac{s}{s_{\sigma}}\right)^{\alpha_{\sigma}(t)} ,\\
    \frac{1}{t-m_{f_1}^2} \to \mathcal{P}_{f_1}(s,t) =&\  \frac{\pi \alpha_{f_1}'\times D_{f_1}(t)}{\Gamma[\alpha_{f_1}(t)]\sin{[\pi \alpha_{f_1}(t)}]} \times \left(\frac{s}{s_{f_1}}\right)^{\alpha_{f_1}(t)-1} ,\\
    \frac{1}{t-m_{f_2}^2} \to \mathcal{P}_{f_2}(s,t) =&\  \frac{\pi \alpha_{f_2}'\times D_{f_2}(t)}{\Gamma[\alpha_{f_2}(t)-1]\sin{[\pi \alpha_{f_2}(t)}]} \times \left(\frac{s}{s_{f_2}}\right)^{\alpha_{f_2}(t)-2} ,
\end{align}
with $s_{\sigma}=s_{f_1}=s_{f_2}=1.0 \:\text{GeV}^2 $. The Regge trajectories $\left(\alpha_{\sigma}(t),\alpha_{f_1}(t), \alpha_{f_2}(t)\right)$ and phases $\left(D_{\sigma}(t),D_{f_1}(t),D_{f_2}(t)\right)$ take the following form: 
\begin{gather}
    \alpha_{\sigma}(t) = \alpha^0_{\sigma} + \alpha_{\sigma}'t = -0.175 + 0.7\:t ,\\
    \alpha_{f_1}(t) = \alpha^0_{f_1} + \alpha_{f_1}'t = 0.95 + 0.028\:t ,\\
    \alpha_{f_2}(t) = \alpha^0_{f_2} + \alpha_{f_2}'t = 0.537 + 0.9\:t ,\\
    D_{\sigma}(t) = \frac{e^{-i\pi\alpha_{\sigma}(t)} + 1}{2},\\
    D_{f_1}(t) = \frac{e^{-i\pi\alpha_{f_1}(t)} - 1}{2},\\
    D_{f_2}(t) = \frac{e^{-i\pi\alpha_{f_2}(t)} + 1}{2}.
\end{gather}

The coupling constants in Eqs.~(\ref{eq:ml_gammaphiphi},\ref{eq:ml_gammaSphi},\ref{eq:ml_gammaAphi}) can be determined by the radiative decays of $\phi$ and $f_1$. 
Using the branching ratios data in PDG~\cite{Workman:2022ynf}: 
\begin{align}
    \text{Br}_{\phi\to\pi^0\gamma} &= (1.32\pm0.05)\times10^{-3}, \cr
    \text{Br}_{\phi\to\eta\gamma} &= (1.301\pm0.025)\times10^{-2}, \cr
    \text{Br}_{\phi\to a_0\gamma} &= (7.6\pm0.6)\times10^{-5}, \cr
    \text{Br}_{\phi\to f_0\gamma} &= (3.22\pm0.19)\times10^{-4}, \cr
    \text{Br}_{f_1\to \gamma\phi} &= (7.4\pm2.6)\times10^{-4}, \nonumber
\end{align}

 we can get the relevant coupling constants: 
 \begin{align*}
     g_{\gamma\pi\phi}&=-0.14, \cr
     g_{\gamma\eta\phi}&=-0.71, \cr
     g_{\gamma a_0\phi}&=-0.77, \cr
     g_{\gamma f_0\phi}&=-2.44, \cr
     g_{\gamma f_1\phi}&=0.17.
 \end{align*}

The coupling constants in Eqs.~(\ref{eq:ml_phiNN},\ref{eq:ml_SNN}) can be determined by Nijmegen potential as: 
\begin{align*}
    g_{\pi NN}&=13.0, \cr 
    g_{\eta NN}&=6.34, \cr
    g_{a_0 NN}&=4.95, \cr
    g_{f_0 NN}&=-0.51. 
\end{align*}

The other coupling constants are considered as fitting parameters within some given ranges. 
For the $\sigma$ exchange, $g_\sigma = g_{\gamma\sigma\phi} g_{\sigma NN} \in (0.5,1.5)$. The selection of the range can be found in Ref.~\cite{Yu:2016zut}.
For the $f_1$ exchange, the value of $g_{f_1NN}$ can be taken from 2.0~\cite{Birkel:1995ct} to 5.8~\cite{Yan:2021tcp}.
So, $g_{f_1} = g_{\gamma f_1\phi} g_{f_1 NN} \in (0.3,1.0)$.
For the $f_2$ exchange, we take $g_{f_2} = g_{\gamma f_2\phi} g_{f_2 NN} \in (0.1,1.0)$.
 
The cutoff parameters $(\Lambda_{\varphi},\Lambda_{S},\Lambda_{\sigma},\Lambda_{f_1},\Lambda_{f_2})$ and the amplitude phases $(\beta_{\varphi},\beta_{S},\beta_{\sigma},\beta_{f_1},\beta_{f_2})$ are the fitting parameters.
\subsection{Proton exchange}
Considering proton exchange in s- and u- channel, from the perspective of gauge invariance, the two amplitudes must be taken together. 
The associated effective Lagrangians can be written as:
\begin{eqnarray}
    \Lag_{\gamma NN} &=& -e \bar{N} (\gamma_\mu - \frac{\kappa_N}{2M_N} \sigma_{\mu\nu}\partial^{\nu}) N A^{\mu}, \\
    \Lag_{\phi NN} &=& -g_{\phi NN} \bar{N} (\gamma_\mu - \frac{\kappa_{\phi NN}}{2M_N} \sigma_{\mu\nu}\partial^{\nu}) N \phi^{\mu}.
\end{eqnarray}

And the amplitudes can be written as:
\begin{gather}
    \mM_{N}^{\mu\nu}(p_1,k_1,p_2,k_2) = (\mathcal{M}_{s-ch,p}^{\mu\nu} + \mathcal{M}_{u-ch,p}^{\mu\nu}) F_p(s,u), \label{eq:Amp:N} \\
	\mathcal{M}_{s-ch,p}^{\mu\nu}(p_1,k_1,p_2,k_2) = \frac{e {g_{\phi NN}}}{s-M_N^2} \left( \gamma^\nu - i\frac{{{\kappa_{\phi NN}}}}{2M_N} \sigma^{\nu\alpha} k_{2\alpha} \right) \cr
	(\rlap{/}{q}_s + M_N) \left( \gamma^\mu + i\frac{{\kappa_p}}{2M_N} \sigma^{\mu\beta} k_{1\beta} \right),
	\cr
	\mathcal{M}_{u-ch,p}^{\mu\nu}(p_1,k_1,p_2,k_2) = \frac{e {g_{\phi NN}}}{u-M_N^2} \left( \gamma^\mu + i\frac{{\kappa_p}}{2M_N} \sigma^{\mu\alpha} k_{1\alpha} \right) \cr (\rlap{/}{q}_u + M_N) \left( \gamma^\nu - i\frac{{\kappa_{\phi NN}}}{2M_N} \sigma^{\nu\beta} k_{2\beta} \right) ,
\end{gather}
where $q_s=k_1+p_1$, $q_u=p_1-k_2$, $u =q_u^2$.
$F_p(s,u)$ is a form factor, taken from Ref.~\cite{Kim:2021adl}:
\begin{gather}
    F_p(s,u) = e^{i\beta_{p}}[F_p(s) + F_p(u) - F_p(s) F_p(u)]^2, \\
    F_p(s) = \frac{\Lambda_p^4}{\Lambda_p^4+(s-M_p^2)^2},\cr
    F_p(u) = \frac{\Lambda_p^4}{\Lambda_p^4+(u-M_p^2)^2}.
\end{gather}

In this part, the coupling constant $\kappa_{N}$ is the anomalous magnetic moment of the nucleon, for proton: $\kappa_{p}=1.79$ .
And $g_{\phi NN}$, $\kappa_{\phi NN}$ can be determind by Nijmegen potential model~\cite{Rijken:1998yy,Stoks:1999bz}. 
There we take $g_{\phi NN}=-1.47$, $\kappa_{\phi NN}=-1.65$~\cite{Kim:2021adl}.
And the cutoff $\Lambda_{p}$ and relative phase $\beta_{p}$ are fitting parameters.
\subsection{\texorpdfstring{$N^*$}{} molecule exchange}

As the $P_c$ states observed by the LHCb collaboration can be well described by the S-wave $\bar{D}\Sigma_c$, $\bar{D}\Sigma^*_c$ and $\bar{D}^* \Sigma_c$ molecular states~\cite{Wu:2010jy,Wu:2010vk,Shen:2016tzq,Lin:2017mtz} with three $\bar D^*\Sigma^*_c$ bound states with $J^P={\frac12}^-, {\frac32}^- \& {\frac52}^-$ expected to exist from heavy quark spin symmetry~\cite{Xiao:2013yca,Liu:2019tjn,Du:2019pij}, 
their hidden strange partners are also expected to exist~\cite{He:2017aps,Lin:2018kcc}.
For the process of $\gamma p \to \phi p$,  the CLAS data~\cite{CLAS:2013jlg,Dey:2014tfa} observed two-peaks around $K^*\Sigma$ and $K^*\Sigma^*$ thresholds, respectively. We expect them to be due to $K^*\Sigma$ and $K^*\Sigma^*$ molecular states, denoted as $N^*(2080)$ and $N^*(2270)$, respectively. Since the energy range is not far away from the $p\phi$ threshold, the S-wave dominant. We consider only $N^*(2080)(3/2^-)$~\cite{Workman:2022ynf} and $N^*(2270)(1/2^-\text{or}\: 3/2^-)$.
For the process $N^*\to K^*\Sigma^{(*)} \to \gamma(\phi) N$ , it should be a triangle diagram, see Appendix~\ref{AppendixA} for details.
For simplicity, we describe these two processes with tree level diagrams. 
And the corresponding coupling constants are used as fitting parameters, and finally compared with the results of the triangle diagrams.

The corresponding effective Lagrangians can be written as~\cite{Oh:2007jd}:
\begin{align}
    \Lag_{RN\gamma}(\textstyle\frac12^\pm) =&\  \frac{ef_1}{2M_N} \bar{N} \Gamma^{(\mp)} \sigma_{\mu\nu} \partial^\nu A^\mu R ,\\
	\Lag_{RN\gamma}(\textstyle\frac32^\pm) =& -\frac{ief_1}{2M_N} \bar{N} \Gamma^{(\pm)}_\nu F^{\mu\nu} R_\mu \nonumber \\
	&- \frac{ef_2}{(2M_N)^2} \partial_\nu \bar{N} \Gamma^{(\pm)} F^{\mu\nu} R_\mu ,\\
	\Lag_{ R N V }(\textstyle\frac12^\pm)
	=&\  \bar{R} \left[
		\pm \frac{g_1 M_V^2}{2M_N(M_R \mp M_N)}\Gamma_\mu^{(\mp)} N \right.\cr
	& +\left. \frac{g_2}{2M_N} \Gamma^{(\mp)} \sigma_{\mu\nu} N \partial^\nu \right] V^{\mu} ,\cr
	\Lag_{ R N V}(\textstyle\frac32^\pm) =&\  \bar{R}_\mu \left[ \frac{ig_1}{2M_N} \Gamma_\nu^{(\pm)} N \pm \frac{g_2}{(2M_N)^2} \Gamma^{(\pm)} \partial_\nu N \right.\cr
	& \mp\left. \frac{g_3}{(2M_N)^2} \Gamma^{(\pm)} N \partial_\nu \right] V^{\mu\nu} ,
\end{align}
where $R$ and $R_\mu$ are the fields for the spin-1/2 and 3/2 resonances, respectively. V and N are the spin-1 vector meson and baryon, respectively. And $V^{\mu\nu}=\partial^{\mu}V^{\nu}-\partial^{\nu}V^{\mu}$,
with 
$ \Gamma_\mu^{(\pm)} =  \left(
		\begin{array}{c}
			\gamma_\mu \gamma_5 \\ \gamma_\mu
		\end{array} 
		\right)$,
$ \Gamma^{(\pm)} = \left(
		\begin{array}{c}
			\gamma_5 \\ \mathbf{1}
		\end{array}
	\right) $.

Then the $N^*$ exchange amplitude can be written as
\begin{widetext}
\begin{align}
    \mM_{s-ch,N^*}^{\mu\nu}\left(1/2^{-}\right) =& F_{R}(s)\left( \frac{g_1 M_V^2}{2M_N(M_R \mp M_N)}\gamma_\nu \gamma_5 + \frac{g_2}{2M_N} i \sigma_{\nu\alpha} k_2^{\alpha} \gamma_5 \right) G^{\frac{1}{2}}(q_s) \frac{ef_1}{2M_N} \gamma_5 (-i) \sigma_{\mu\beta} k_1^{\beta} ,\\
    \mM_{s-ch,N^*}^{\mu\nu}\left(3/2^-\right) =& F_{R}(s)\left( \frac{g_1}{2M_N}\gamma_\rho + \frac{(g_2 p_2-g_3 k_2)_{\rho}}{4M_N^2} \right) (k_2^{\alpha} g^{\rho\mu}- k_2^{\rho} g^{\alpha\mu})  G^{\frac{3}{2}}_{\alpha\beta}(q_s) \cr
    & \times \left(\frac{ef_1}{2M_N} \gamma_{\sigma} + \frac{ef_2}{4M_N^2} (p_1)_{\sigma} \right) (k_1^{\beta} g^{\sigma\mu}- k_1^{\sigma} g^{\beta\mu}),
\end{align}
\end{widetext}
where
\begin{align}
    G^{\frac{1}{2}}(p) =&\  \frac{i (\rlap{/}{p} + M_R) }{s-M_R^2+i M_R \Gamma_R} ,\\
    G^{\frac{3}{2}}_{\mu\nu}(p) =&\  \frac{i (\rlap{/}{p} + M_R) }{s-M_R^2+i M_R \Gamma_R} \left( -g_{\mu\nu} + \frac{1}{3}  \gamma_{\mu}\gamma_{\nu} \right.\cr
    & \left.+ \frac{2}{3} \frac{p_{\mu}p_{\nu}}{M_R^2} +\frac{1}{3M_R}(\gamma_\mu p_\nu-\gamma_\nu p_\nu) \right).
\end{align}

And form factor $F_{R}(s)$ takes following form:
\begin{align}
    F_{R}(s) = e^{i\beta_{R}}e^{-\frac{(s-m_{R}^2)^2}{\Lambda_{R}^4}} .
\end{align}

For the $N^*(2080)$, $J^P =3/2^-$, there are these parameters need to be fitted:
\begin{gather*}
    h_2(f_2/f_1),\: g_1,\: g_2,\: g_3,\: \beta_{N^*_1},\: \Lambda_{N^*_1},\: \Gamma_{N^*_1}.
\end{gather*}

For the $N^*(2270)$, $J^P =3/2^-(1/2^-)$, the fitted parameters are: 
\begin{gather*}
    h'_2(f_2'/f_1'),\: g'_1,\: g'_2,\: g'_3,\: \beta_{N^*_2},\: \Lambda_{N^*_2},\: \Gamma_{N^*_2},\cr
    ( g'_1/f_1',\: g'_2/f_1',\: \beta_{N^*_2},\: \Lambda_{N^*_2},\: \Gamma_{N^*_2}).
\end{gather*}

\section{Results and discussions} \label{sec:result}
According to the $K\bar{K}$  decay mode of the $\phi$, the CLAS Collaboration divided the results of $\gamma p \to \phi p$ into charged-$(\phi \to K^+ K^-)$ and neutral-$(\phi \to K^0_S K^0_L)$ mode.
Considering the contamination of the process $\gamma p \to K^+ \Lambda(1520)/\Lambda(1800) \to p K^+ K^-$ on the charged mode, although relevant cuts have been done in the experiment, we only use the neutral-mode to fit.

\begin{figure*}[htbp]
	\centering
	\includegraphics[width=0.9\textwidth]{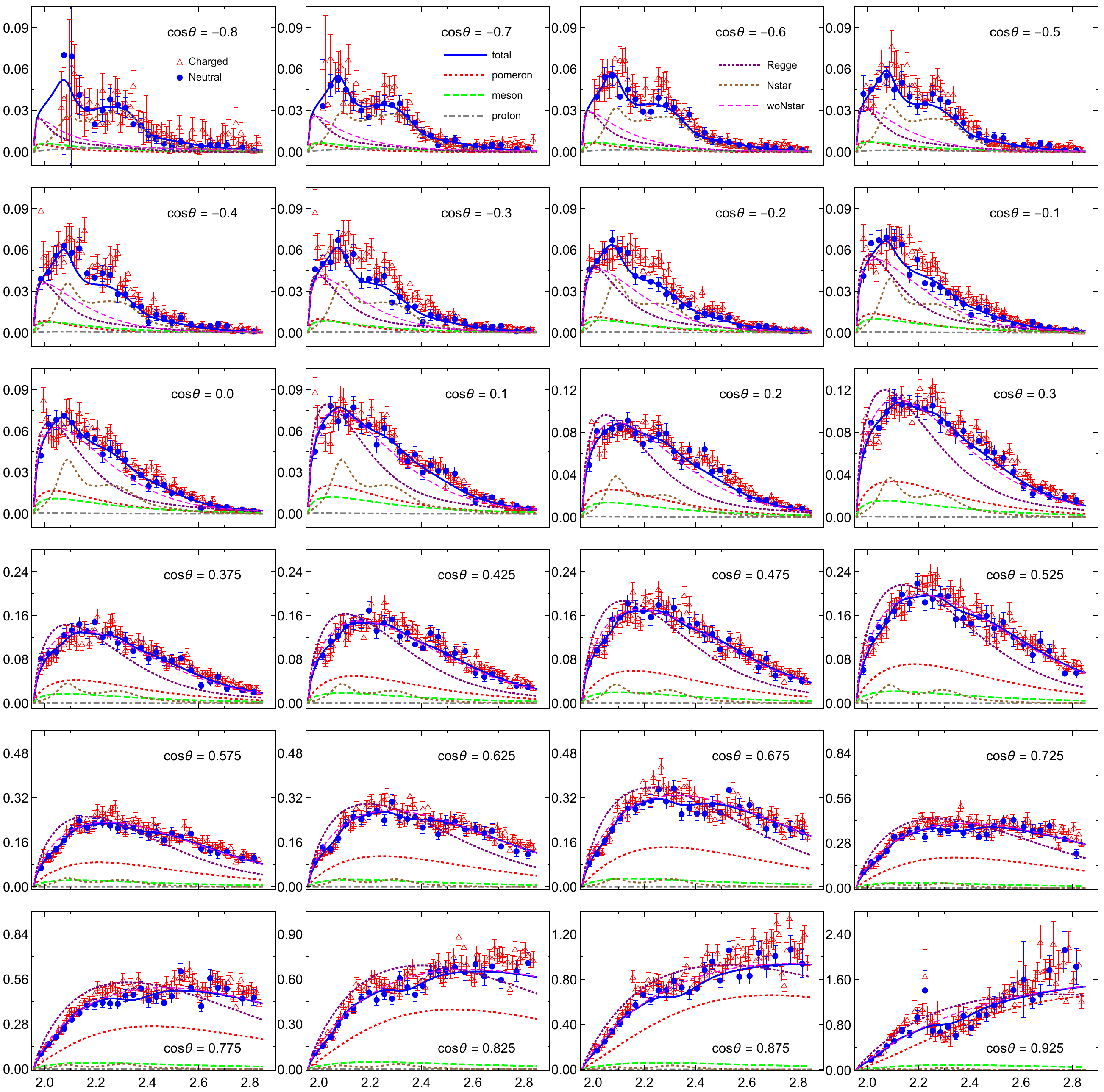}
	\caption{\label{fig:dsigmadcs} Differential cross sections $d\sigma/d W(\mu\text{b}/\text{GeV})$ at different  $\cos{\theta}$. 
	The blue solid line stands for the total contribution. 
	The red dotted line, green dashed line, gray dash-dotted line, purple dotted line and brown dotted line are the contributions from pomeron exchange , pseudo-scalar and scalar mesons $(\pi,\eta,a_0,f_0)$ exchange, s- and u-channel proton exchange, Reggeized mesons $(\sigma,f_1,f_2)$ exchange and two $N^*$ molecules exchange, respectively.
	The magenta dashed line is the full contribution without $N^*$.
	The experimental data are taken from Ref.~\cite{Dey:2014tfa}, where the red triangle and blue circle represent the charged-mode and the neutral-mode, respectively.}
\end{figure*}

\begin{figure*}[htbp]
	\centering
	\includegraphics[width=0.9\textwidth]{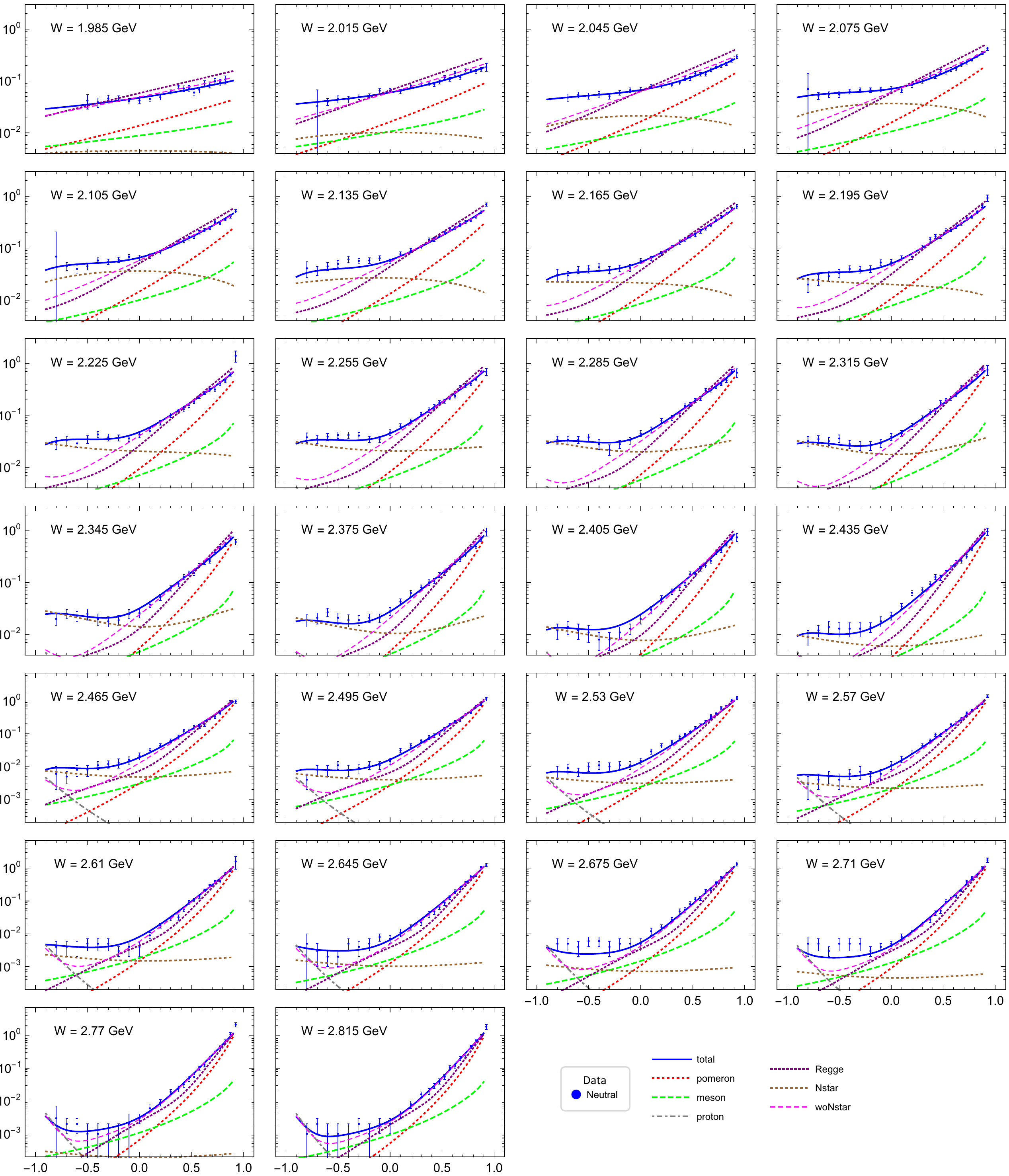}
	\caption{\label{fig:dsigmadw}Differential cross sections $d\sigma/d \cos{\theta}(\mu\text{b})$ at different c.m. energies $W$.
	The marks are the same as in Fig.~\ref{fig:dsigmadcs}.}
\end{figure*}

\begin{figure*}[htbp]
	\centering
	\includegraphics[width=0.9\textwidth]{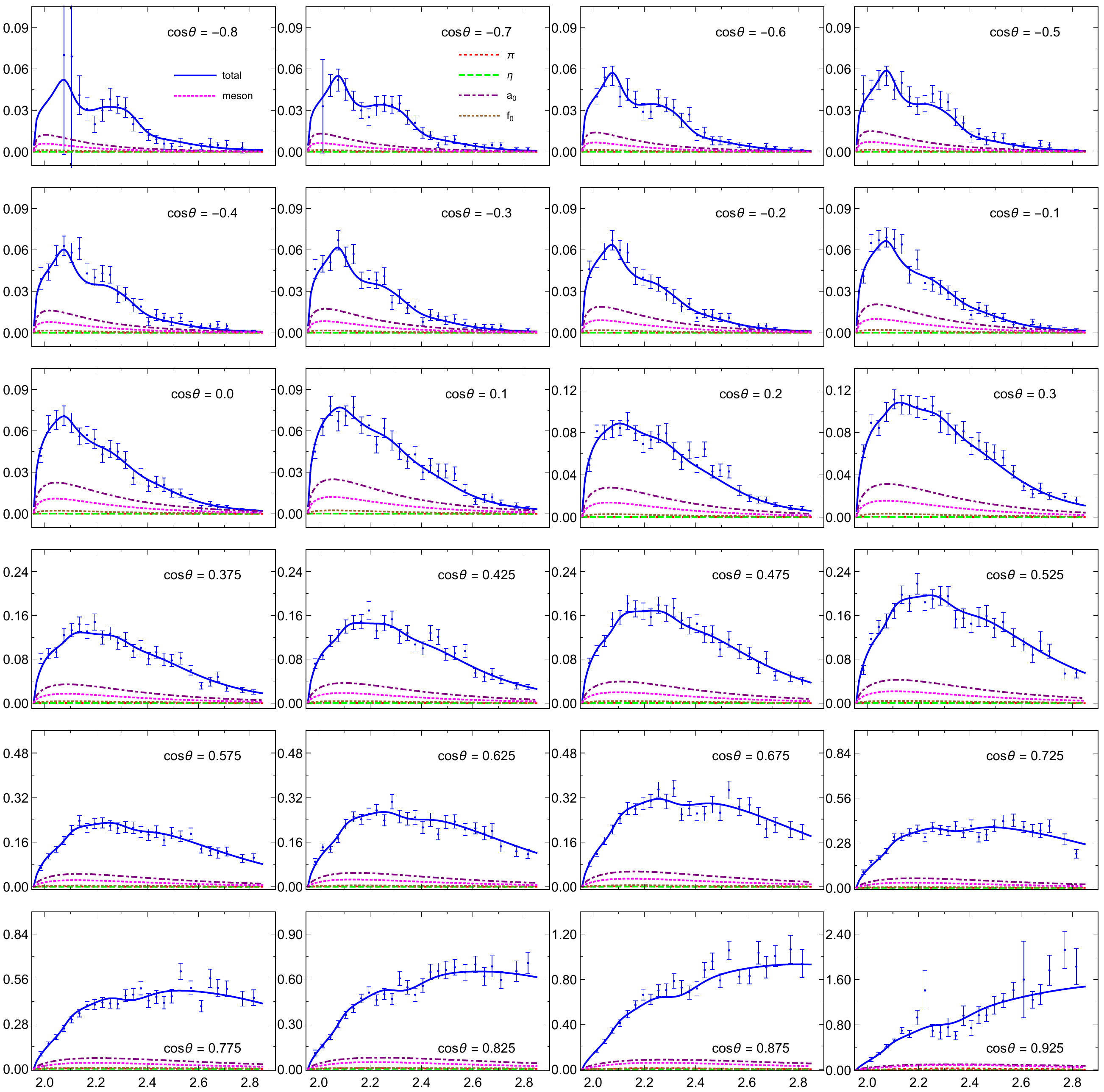}
	\caption{\label{fig:mesondcsx}Mesons contribution to the differential cross sections $d\sigma/d W (\mu\text{b}/\text{GeV})$ at different $\cos\theta$.}
\end{figure*}

\begin{figure*}[htbp]
	\centering
	\includegraphics[width=0.9\textwidth]{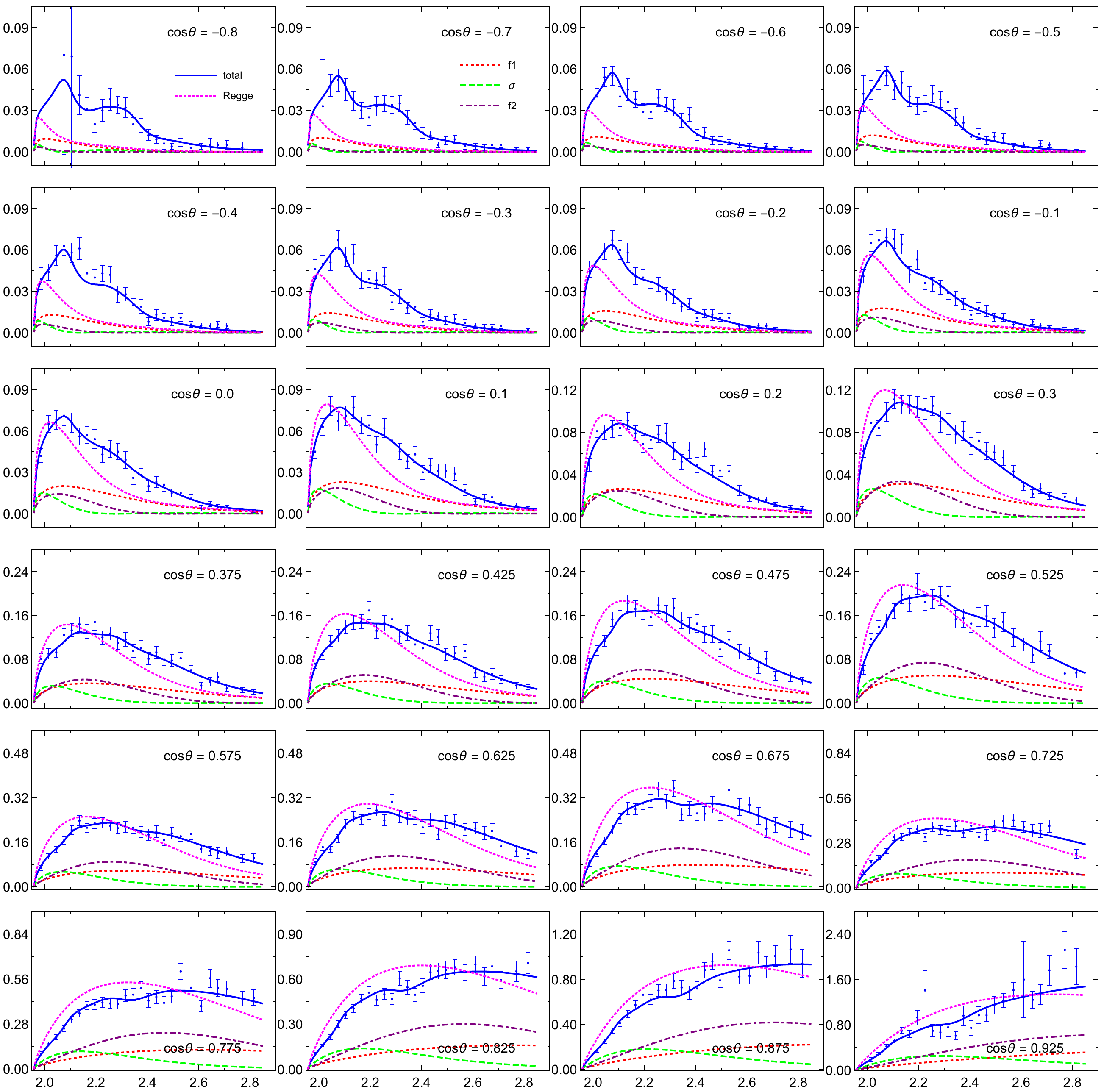}
	\caption{\label{fig:reggedcsx}Regged mesons contribution to the differential cross sections$d\sigma/d W (\mu\text{b}/\text{GeV})$ at different $\cos\theta$.}
\end{figure*}

\begin{figure}[htbp]
	\centering
	\includegraphics[width=0.45\textwidth]{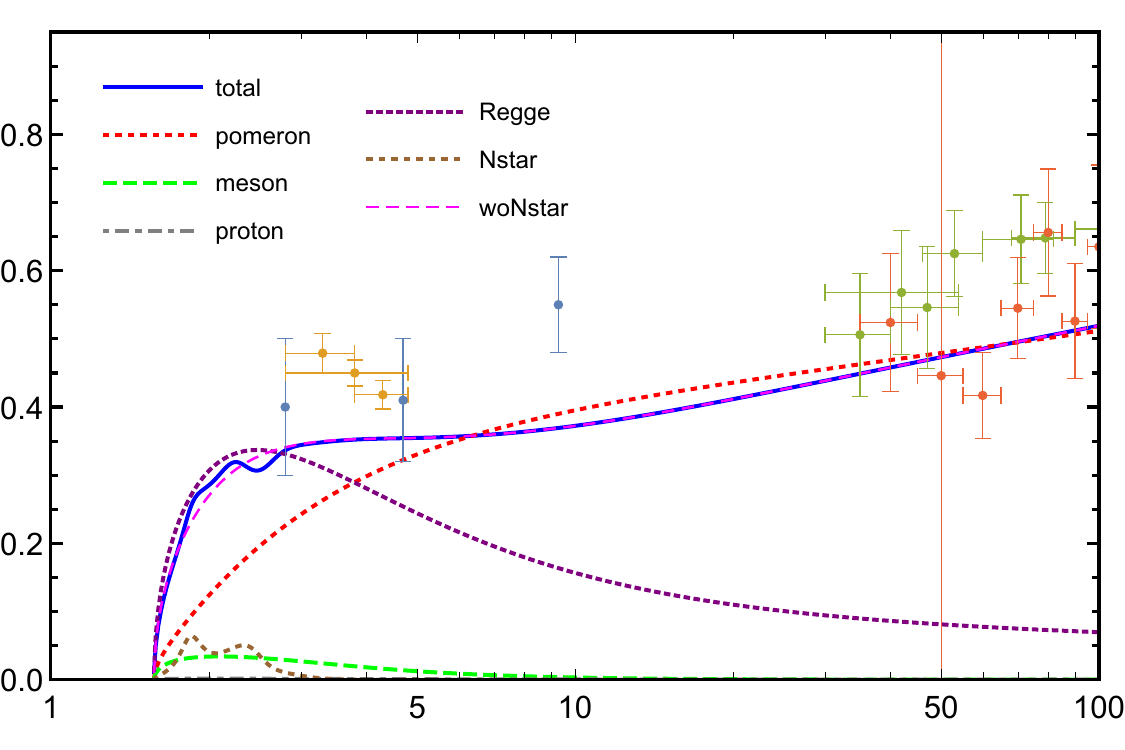}
	\caption{\label{fig:totalsigmax}Total cross section $\sigma(\mu\text{b})$ as a function of the photon energy
	in the laboratory frame $E^{\text{lab}}_\gamma(\text{GeV})$.
	Experimental data comes from Refs.~\cite{Ballam:1972eq,Egloff:1979mg,Barber:1981fj,Busenitz:1989gq}:}
\end{figure}

By using the MINUIT algorithm~\cite{James:1975dr,iminuit,iminuit.jl} to minimize the $\chi^2$ function, we fitted the experimental data of the differential cross-section $d\sigma/d\cos\theta$ in Eq.~({\ref{eq:dsigma}}) and obtained the best fitting result $\chi^2/dof=0.87$.
When $J^P(N^*(2270))=1/2^-$, we do not get good fitting results, so we only show $J^P=3/2^-$ results.
The corresponding parameters and results are shown in Table~\ref{table:fit_parameters} and Figs.~(\ref{fig:dsigmadcs},\ref{fig:dsigmadw},\ref{fig:mesondcsx},\ref{fig:reggedcsx},\ref{fig:totalsigmax}).

\begin{table*}[htbp]
	\caption{\label{table:fit_parameters}The model parameters.}
	\renewcommand\arraystretch{2.0}
    \centering
    \begin{tabular}{|p{0.07\textwidth}<{\centering}|p{0.13\textwidth}<{\centering}|p{0.07\textwidth}<{\centering}|p{0.13\textwidth}<{\centering}|p{0.09\textwidth}<{\centering}|p{0.13\textwidth}<{\centering}|p{0.09\textwidth}<{\centering}|p{0.13\textwidth}<{\centering}|}
    \hline 
    \multicolumn{2}{|c|}{phases} & \multicolumn{2}{c|}{cutoffs~(GeV)} & \multicolumn{4}{c|}{other constants} \\
    \hline 
    $\displaystyle \beta _{ps}$ & $ 2.699\pm0.027 $ & $\displaystyle \Lambda _{ps}$ & $ 0.36\pm0.07$ & $\displaystyle g_{\sigma }$ & $0.540\pm0.029$ & $\displaystyle \Gamma _{N_{2}^{*}}\left(\text{GeV}\right)$ & $0.23\pm0.04$ \\
    \hline 
    $\displaystyle \beta _{s}$ & $3.255\pm0.033$ & $\displaystyle \Lambda _{s}$ & $0.751\pm0.028$ & $\displaystyle g_{f_{1}}$ & $0.86\pm0.17$ & $\displaystyle g'_{1}$ & $2.5\pm0.5$ \\
    \hline 
     $\displaystyle \beta _{p}$ & $5.76\pm0.06$ & $\displaystyle \Lambda _{p}$ & $0.766\pm0.021$ & $\displaystyle g_{f_{2}}$ & $0.25\pm0.04$ & $\displaystyle g'_{2}$ & $-2.0\pm0.6$ \\
    \hline 
    $\displaystyle \beta _{\sigma }$ & $2.830\pm0.028$ & $\displaystyle \Lambda _{\sigma }$ & $2.00\pm0.02$ & $\displaystyle \Gamma _{N_{1}^{*}}\left(\text{GeV}\right)$ & $0.099\pm0.011$ & $\displaystyle g'_{3}$ & $2.5\pm0.6$ \\
    \hline 
    $\displaystyle \beta _{f_{1}}$ & $0.419\pm0.004$ & $\displaystyle \Lambda _{f_{1}}$ & $1.47\pm0.06$ & $\displaystyle g_{1}$ & $0.0395\pm0.0024$ & $\displaystyle h'_{2}$ & $-0.90\pm0.05$ \\
    \hline 
    $\displaystyle \beta _{f_{2}}$ & $3.027\pm0.030$ & $\displaystyle \Lambda _{f_{2}}$ & $1.96\pm0.15$ & $\displaystyle g_{2}$ & $0.095\pm0.007$ &  &  \\
    \hline 
    $\displaystyle \beta _{N_{1}^{*}}$ & $2.733\pm0.027$ & $\displaystyle \Lambda _{N_{1}^{*}}$ & $1.432\pm0.022$ & $\displaystyle g_{3}$ & $0.1937\pm0.0034$ &  &  \\
    \hline 
    $\displaystyle \beta _{N_{2}^{*}}$ & $1.001\pm0.010$ & $\displaystyle \Lambda _{N_{2}^{*}}$ & $0.752\pm0.028$ & $\displaystyle h_{2}$ & $-9.5\pm0.5$ &  &  \\
    \hline
    \end{tabular}
\end{table*}


From the fitting result, we can find the proton exchange has little contribution in all regions.

For the Pomeron exchange, it contributes little in the backward angle area. 
Its contribution starts to increase as $\cos{\theta}$ increases. 
It is the main contributor to the differential cross section as $\cos{\theta}\sim 1$, and increases with the center-of-mass energies in this region.
In other angles, its contribution tends to increase first and then decrease.

For meson exchange of $(\pi,\eta,a_0,f_0)$, the overall contribution to the differential scattering cross section is relatively small, and also increases first and then decreases with the center-of-mass energies.
Among them, The contribution of pseudo-scalar mesons are much smaller than which from scalar mesons, and the contributions of $\eta$ and $f_0$ are much smaller than those of $\pi$ and $a_0$ respectively.

For the Reggeized parameterized mesons $(\sigma,f_1,f_2)$ exchange, they have little effect in the backward angle except near the threshold.
Their contribution to the cross-section becomes larger as the angle increases.
And their change trend with the center-of-mass system energy is the same as that of meson exchange, there is a peak, and the peak position and height increase with the increase of the angle.

For $N^*$ molecules exchange, the contribution in the backward corner region is the most dominant, and the contribution in other regions is smaller.

Considering the behavior of the total cross-section, in the high-energy region, besides the largest contribution of the Pomeron, only Reggeized parameterized mesons contribute, and due to their relative phase, the exchange of reggeized mesons has little effect on the total cross-section.

From fitting results about $N^*$ molecules exchange:
\begin{gather*}
    \Gamma_{N^*_1} = 0.099\pm0.011\:\text{GeV},\quad
    \Gamma_{N^*_2} = 0.23\pm0.04\:\text{GeV}, \cr
    g_1=0.0395\pm0.0024,\quad g_2=0.095\pm0.007,\cr
    g_3=0.1937\pm0.0034, \quad f_2/f_1=-9.5\pm0.5 ,\cr
    g'_1=2.5\pm0.5,\quad g'_2=-2.0\pm0.6,\cr
    g'_3=2.5\pm0.6, \quad f'_2/f'_1=-0.90\pm0.05 ,
\end{gather*}
we can calculate the following properties:
\begin{gather*}
    A_{3/2}/A_{1/2}=1.60\pm0.13,\quad
    A'_{3/2}/A'_{1/2}=-1.12\pm0.25, \\
    \Gamma_{N^*_1\to p\phi} \times \Gamma_{N^*_1\to \gamma p} = (8.2\pm2.6)\times10^{-4} \:\text{MeV}^2 ,\\
    \Gamma_{N^*_2\to p\phi} \times \Gamma_{N^*_2\to \gamma p} = (6\pm6)\times10^{-3} \:\text{MeV}^2 .
\end{gather*}

The above results are roughly consistent with the calculation from the molecular state triangle diagram in Appendix~\ref{AppendixA} at small cutoffs $\Lambda_0\sim(0.6,0.8)\:\text{GeV}$ and $\Lambda_1\sim(0.8,1.0)\:\text{GeV}$.

Then, we discuss other mechanisms that may affect this process.

Firstly, for the charged-mode, we need to consider the final state of the three-body decay.
The corresponding process is $\gamma p \to \phi p, K^+\Lambda(1520), K^+\Lambda(1800) \to p K^+ K^-$.
After adding the corresponding hard cuts of $M_{p K^-}$ according to the experimental analysis, we can fit the charged-mode data and the neutral-mode data at the same time.

Secondly, the thresholds of $K^+\Lambda(1520)$ and $p\phi$ are very close.
So there may be coupled-channel effects~\cite{Ryu:2012tw}. 
Then we can fit the experiment data of $K^+\Lambda(1520)$ and $p\phi$ at the same time.

In summary, by fitting the experimental data of CLAS in 2014, we carefully analyzed the process of $\gamma p\to \phi p$. It is found that the data can be fitted well by using $N^*(2080)$ and $N^*(2270)$ instead of previous $N^*(2000,5/2^+)$ and $N^*(2300,1/2^+)$ for the s-channel $N^*$ exchange together with other background terms.
The fitted coupling constants of these $N^*$ molecular states to $p\phi$ and $\gamma p$ are consistent with the results directly calculated from the relevant hadronic triangle diagrams of the molecular picture.
The new solution gives a natural explanation of the two $N^*$ peaks in  the process of $\gamma p\to \phi p$,
meanwhile a further support of the existence of the strange molecular partners of $P_c$ states. 
%
\section*{Acknowledgments}
We thank useful discussions and valuable comments from Feng-Kun Guo and Jia-Jun Wu. 
This work is supported by the NSFC and the Deutsche Forschungsgemeinschaft (DFG, German Research
Foundation) through the funds provided to the Sino-German Collaborative
Research Center TRR110 “Symmetries and the Emergence of Structure in QCD”
(NSFC Grant No. 12070131001, DFG Project-ID 196253076 - TRR 110), by the NSFC 
Grant No.11835015, No.12047503, and by the Chinese Academy of Sciences (CAS) under Grant No.XDB34030000.

\begin{appendix}

\section{\texorpdfstring{$N^*$}{} hadronic molecules }\label{AppendixA}

For the $N^*(2270)$ as the $K^* \Sigma^*$ molecular, it should be mentioned that the sets of spin and parity for $(K^*,\Sigma^*)$ is $(1^-,3/2^+)$. 
Thus the $N^*$ states of spin-parity $(1/2^-, 3/2^-, 5/2^-)$ may be considered as S-wave bound states of $K^*\Sigma^*$. 
Subject to the Lorentz covariant orbital-spin scheme , the S-wave couplings for the $N^*$ with $J^P = 1/2^-, 3/2^-$ with the meson–baryon pairs of interest are given by:
\begin{eqnarray}
	\Lag_{K^{*} \Sigma^{*} N^{*}(1/2^-)} &=& g_{K^{*} \Sigma^{*} N^{*}}^{1/2^-} \bar{\Sigma}^*_\mu N^* K^{* \mu}  ,\label{eq:lag1} \\[4pt]
	\Lag_{K^{*} \Sigma^{*} N^{*}(3/2^-)} &=& g_{K^{*} \Sigma^{*} N^{*}}^{3/2^-} \bar{\Sigma}^{* \mu} \gamma_5 \tilde{\gamma}^{\nu} N^*_{\mu} K^{*}_{\nu} ,\label{eq:lag2}
\end{eqnarray}{\tiny }
where $ \tilde{\gamma}^{\nu} = (g_{\mu \nu} - \frac{p_\mu p_\nu}{p^2}) \gamma^\mu $ with $ p_\mu $ the momentum of initial $N^*$ state.
And two S-wave coupling constants $g_{K^{*} \Sigma^{*} N^{*}}^{1/2^-}$ and $g_{K^{*} \Sigma^{*} N^{*}}^{3/2^-}$ can be estimated by the Weinberg compositeness criterion~\cite{Weinberg:1965zz,Baru:2003qq,Guo:2017jvc}:
\begin{eqnarray}
	g_{K^{*} \Sigma^{*} N^{*}}^{1/2^-} &=& \sqrt{\frac{2\pi m_2\sqrt{2 \epsilon}}{\mu^{3/2}}} ,\\[4pt]
	g_{K^{*} \Sigma^{*} N^{*}}^{3/2^-} &=& \sqrt{\frac{12\pi m_2\sqrt{2 \epsilon}}{5\mu^{3/2}}},
\end{eqnarray}{\tiny }
where $\mu=m_1 m_2/(m_1+m_2)$ is the reduced mass of the bound particles, $m_1$ and $m_2$ denote the masses of $\Sigma^*$ and $K^*$, respectively, and $\epsilon= m_1 + m_2 - M $ is the binding energy.

Similarly, for the $N^*(2080)(3/2^-)$ as the $K^* \Sigma$ molecular, the S-wave coupling can be writted:
\begin{eqnarray}
	\Lag_{K^{*} \Sigma N^{*}(3/2^-)} &=& g_{K^{*} \Sigma N^{*}}^{3/2^-} \bar{\Sigma} N^*_{\mu} K^{*\mu} ,\label{eq:lag3}\\[4pt]
	g_{K^{*} \Sigma N^{*}}^{3/2^-} &=& \sqrt{\frac{4\pi m_2\sqrt{2 \epsilon}}{\mu^{3/2}}}.
\end{eqnarray}{\tiny }

\begin{figure}[htbp]
	\begin{center}
		\includegraphics[width=0.4\textwidth]{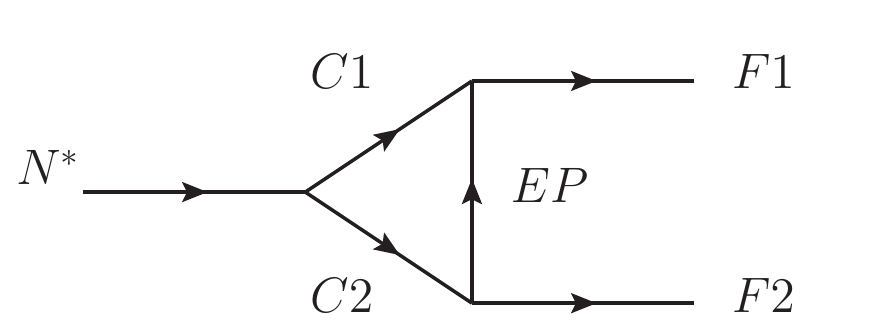}
		\caption{The triangle diagram for the two-body decays of the exotic $N^*$s in the $K^*\Sigma^*$ and $K^*\Sigma$ molecular pictures, where $C1$, $C2$ denote the constituent particles of the composite system $K^*\Sigma^*$ or $K^*\Sigma$, $F1$, $F2$ denote the final states, $EP$ denotes the exchanged particles.\label{Fig:triangle}}
	\end{center}
\end{figure}

As shown in Fig.~(\ref{Fig:triangle}), we calculate the partial decay width of the molecular state by calculating this triangle diagram, the process considered in Table~\ref{Tab:modes}. Therefore we also need the following effective Lagrangians~\cite{Lin:2017mtz}:
\begin{table}[htpb]
	\centering
	\caption{\label{Tab:modes}Used decay channels of $N^*$.}
    \begin{tabular}{c|c|c}
		\Xhline{0.5pt}
		\thead{Initial state} & \thead{Final states} & \thead{Exchanged particles} \\
		\Xhline{0.5pt}
		\multirow{2}*{$N^*(2270)(K^* \Sigma^*)$} & $p\phi$ & $K$, $K^*$ \\
		\Xcline{2-3}{0.4pt}
		& $\gamma p$ & $K$ \\
		\Xhline{0.5pt}
		\multirow{2}*{$N^*(2080)(K^* \Sigma)$} & $p\phi $ & $K$, $K^*$ \\
		\Xcline{2-3}{0.4pt}
		& $\gamma p$ & $K$ \\
		\Xhline{0.5pt}
	\end{tabular}
\end{table}
\begin{eqnarray*}
	\Lag_{V_1 V_2 P} &=& -g_{V_1 V_2 P} \ \varepsilon^{\mu \nu \alpha \beta} \left(\partial_{\mu} V_{1 \nu}  \partial_{\alpha} V_{2 \beta}\right) P,\\[4pt]
	\Lag_{V_1 V_2 V_3} &=& -i g_{V_1 V_2 V_3} \Bigl\{ V_1^{\mu}\left(\partial_{\mu} V_2^{\nu} V_{3 \nu} - V_2^{\nu} \partial_{\mu} V_{3 \nu}\right) \nonumber\\[4pt]
	&& + \left(\partial_{\mu} V_{1 \nu} V_2^{\nu} - V_{1 \nu} \partial_{\mu} V_2^{\nu} \right) V_3^{\mu} \nonumber \\
	&& + V_2^{\mu}\left(V_1^{\nu} \partial_{\mu} V_{3 \nu} - \partial_{\mu} V_{1 \nu} V_3^{\nu}\right) \Bigr\},\\[4pt]
	\Lag_{P B B^*} &=& g_{P B B^*} \bar{B}^{* \mu} \partial_{\mu} P B, \label{eq:PBB*} \\[4pt]
	\Lag_{V B B^*} &=& -i g_{V B B^*} \bar{B}^{* \mu} \gamma^{\nu} \gamma_5 [\partial_{\mu} V_{\nu} - \partial_{\nu} V_{\mu}] B  \label{eq:VBB*} ,\\[4pt]
	\Lag_{P B_1 B_2} &=& -i g_{P B_1 B_2} \bar{B}_1 \gamma_5 B_2 P ,\\[4pt]
	\Lag_{V B_1 B_2} &=& g_{P B_1 B_2} \bar{B}_1 \gamma_{\mu} V^{\mu} B_2 ,
\end{eqnarray*}
where $V_1 V_2 P$ denotes $K^* K \phi$ or $K^* K \gamma$, $V_1 V_2 V_3$ denotes $K^* K^* \phi$ or $K^* K^* \gamma$, $P B B^*$ denotes $K p \Sigma^*$, $V B B^*$ denotes $K^* p \Sigma^*$, $P B_1 B_2$ denotes $K p \Sigma$, $V B_1 B_2$ denotes $K^* p \Sigma$.

Here we list the exact values of the coupling constants involved in our calculations in the Table~\ref{Tab:coupling_constants}.
Where $g_{\gamma K^*K}$ is derived from the radiative decay of $K^*$ in PDG~\cite{Workman:2022ynf}, and the other constants are derived from the SU(3) flavor symmetry~\cite{deSwart:1963pdg,Ronchen:2012eg}. 
\begin{table}[htpb]
	\centering
	\caption{\label{Tab:coupling_constants}The coupling constants used in our calculation.}
	\begin{tabular}{*{4}{c}}
		\Xhline{0.5pt}
		\thead{$g_{KN \Sigma^*}$ \\ $(\mathrm{GeV}^{-1})$} & \thead{$g_{K^*N \Sigma^*}$ \\ $(\mathrm{GeV}^{-1})$} & $g_{K^*K\phi}$ & $g_{K^*K^*\phi}$ \\
		\Xhline{0.5pt}
		$6.202$ & $8.444$ & $9.077$ & $4.271$  \\
		\Xhline{0.5pt}
		\thead{$g_{KN \Sigma}$} & $g_{K^*N \Sigma} $ & \thead{$g_{K_0^*K_0 \gamma}$ \\ $(\mathrm{GeV}^{-1})$} & \thead{$g_{K_c^*K_c \gamma}$ \\ $(\mathrm{GeV}^{-1})$} \\
		\Xhline{0.5pt}
		$ 2.7 $ & $ -3.25 $ &  $ -0.385 $ &  $ 0.253 $ \\
		\Xhline{0.5pt}
		\end{tabular}
\end{table}
	
In order to make the calculation reasonable, we add two form factor. 
The following Gaussian regulator $f_1$ is adopted to suppress short-distance contributions, and the monopole form factor $f_2$ is introduced to suppress the off-shell contributions for the exchanged particles.
\begin{eqnarray}
	f_1(\bm{p}^2 /\Lambda_0^2) = {\rm{exp}}(-\bm{p}^2 /\Lambda_0^2),
	\label{eq:regualtor}\\[4pt]
	f_2(q^2) = \frac{\Lambda_1^4}{(m^2 - q^2)^2 + \Lambda_1^4},
	\label{eq:ff}
\end{eqnarray}
where $\bm p$ is the spatial part of the momentum of $K^*$ and $\Sigma^*$ in the rest frame of the $N^*$ state, $m$ is the mass of the exchanged particle and $q$ is the corresponding momentum, $\Lambda_0$ and $\Lambda_1$ are an ultraviolet cut-off and off-shell cutoff, respectively.
The cutoff $\Lambda_0$ denotes a hard momentum scale which suppresses the contribution of the two constituents at short distances $\sim 1/\Lambda_0$. 
There is no universal criterion for choosing the cut-off, but as a general rule the value of $\Lambda_0$ should be much larger than the typical momentum in the bound state, given by $\sqrt{2\mu\epsilon}$.
And it should also not be too large since we have neglected all other degrees of freedom, except for the two constituents, which would play a role at short distances. 
Here we range $\Lambda_0$ from $0.6 \ \mathrm{GeV}$ to $1.4\ \mathrm{GeV}$.
The cut-off $\Lambda_1$ for the off-shell form factor varies for the different system, and we will vary it in the range of $0.8\ \mathrm{GeV}$ to $2.0\ \mathrm{GeV}$.

\begin{figure}[htbp]
	\centering
	\includegraphics[width=0.4\textwidth]{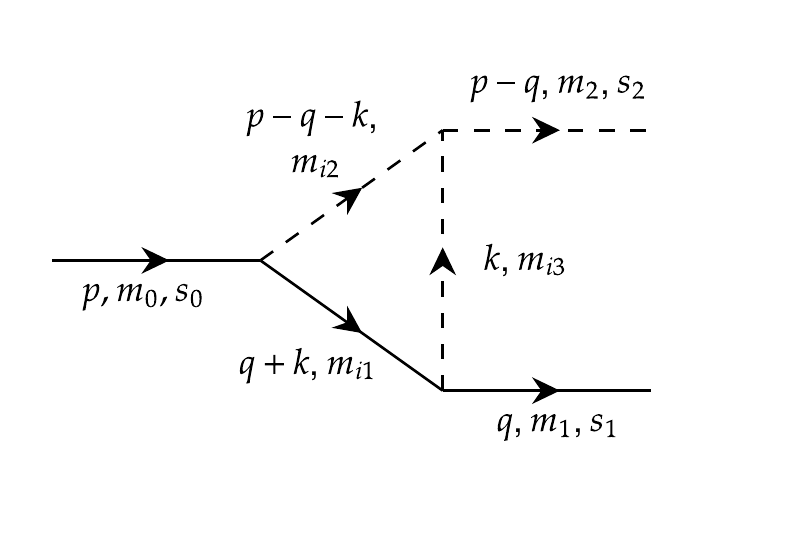}
	\caption{The particle's momentums of the triangle diagram.\label{fig:triangle_momentums}}
\end{figure}

So, following the momentums notation in Fig.~\ref{fig:triangle_momentums}, the corresponding amplitude can be written as: 

\begin{widetext}
\begin{eqnarray}
	\mathcal{M}_i &=& g_0 g_1 g_2 \int_{-\infty}^{\infty} \frac{{\rm d}^4 k}{(2\pi)^4}
	{\rm{exp}}	\left\{\frac{-(\bm{q}+\bm{k})^2}{\Lambda_0^2}\right\} \frac{\Lambda_1^4}{(m_{i3}^2 - k^2)^2 + \Lambda_1^4} \notag\\
	&& \times \frac{A_i(k)}{[(q+k)^2 - m_{i1}^2 + i\epsilon_{i1}][(p-q-k)^2 - m_{i2}^2 + i\epsilon_{i2}](k^2 - m_{i3}^2 + i\epsilon_{i3})},
\end{eqnarray}
\end{widetext}

\begin{widetext}
\begin{eqnarray}
	A^{1/2}_{N^*(2270),K} &=& \bar{u}(q,s_1) (\slashed{q}+\slashed{k} + m_{i1}) \mP_{3/2^+}^{\mu \nu}(q+k,m_{i1}) u(p,s_0) \notag \\
	&& \times k_{\mu} \tilde{g}_{\nu \sigma}^{i2} \epsilon^{\lambda \sigma \alpha \beta} (p-q-k)_{\lambda} (p-q)_{\alpha} \varepsilon^*_{\beta}(p-q,s_2),\\[4pt]
	A^{1/2}_{N^*(2270),K^*} &=& \bar{u}(q,s_1) \gamma_5 \gamma^{\nu} (\slashed{q}+\slashed{k} + m_{i1}) \mP_{3/2^+}^{\mu \alpha}(q+k,m_{i1}) u(p,s_0) \Big\{ k_\mu (p-q+k)^\lambda \tilde{g}_{\alpha \lambda}^{i2} \tilde{g}_{\nu}^{i3 \beta}\notag \\
	&& + k_\mu (p-q-2k)^\beta \tilde{g}_{\alpha \lambda}^{i2} \tilde{g}_{\nu}^{i3 \lambda} 
	+ k_\mu (k-2p+2q)_\lambda \tilde{g}_{\alpha }^{i2 \beta} \tilde{g}_{\nu}^{i3 \lambda}
	- (\mu \leftrightarrow \nu)\Big\} \varepsilon^*_{\beta}(p-q,s_2),\\[4pt]
	A^{3/2}_{N^*(2270),K} &=& \bar{u}(q,s_1) (\slashed{q}+\slashed{k} + m_{i1}) \mP_{3/2^+}^{\mu \nu}(q+k,m_{i1}) \gamma_{5} \tilde{\gamma}^{\rho} u_{\nu}(p,s_0) \notag \\
	&&  k_{\mu} \tilde{g}_{\rho \sigma}^{i2} \epsilon^{\lambda \sigma \alpha \beta} (p-q-k)_{\lambda} (p-q)_{\alpha} \varepsilon^*_{\beta}(p-q,s_2),\\[4pt]
	A^{3/2}_{N^*(2270),K^*} &=& \bar{u}(q,s_1) \gamma_5 \gamma^{\nu} (\slashed{q}+\slashed{k} + m_{i1}) \mP_{3/2^+}^{\mu \alpha}(q+k,m_{i1}) \gamma_{5} \tilde{\gamma}^{\beta} u_{\alpha}(p,s_0) \Big\{k_\mu (p-q+k)^\lambda \tilde{g}_{\beta \lambda}^{i2} \tilde{g}_{\nu}^{i3 \sigma} \notag \\
	&& + k_\mu (p-q-2k)^\sigma \tilde{g}_{\beta \lambda}^{i2} \tilde{g}_{\nu}^{i3 \lambda} 
	+ k_\mu (k-2p+2q)_\lambda \tilde{g}_{\beta }^{i2 \sigma} \tilde{g}_{\nu}^{i3 \lambda}
	- (\mu \leftrightarrow \nu)\Big\} \varepsilon^*_{\sigma}(p-q,s_2),\\[4pt]
	A^{3/2}_{N^*(2080),K} &=& \bar{u}(q,s_1) \gamma_5 (\slashed{q}+\slashed{k} + m_{i1}) u^{\rho}(p,s_0) \tilde{g}_{\nu \rho}^{i2} \epsilon^{\mu \nu \alpha \beta} (p-q-k)_{\mu} (p-q)_{\alpha} \varepsilon^*_{\beta}(p-q,s_2),\\[4pt]
	A^{3/2}_{N^*(2080),K^*} &=& \bar{u}(q,s_1) \gamma_\mu (\slashed{q}+\slashed{k} + m_{i1}) u_{\nu}(p,s_0) \Big\{ (p-q+k)_\alpha \tilde{g}^{\nu \alpha}_{i2} \tilde{g}^{\mu \beta}_{i3} \notag \\
	&& + (p-q-2k)^\beta \tilde{g}^{\nu}_{i2 \alpha} \tilde{g}^{\mu \alpha}_{i3} + (k-2p+2q)_\alpha \tilde{g}^{\nu \beta}_{i2} \tilde{g}^{\mu \alpha}_{i3}\Big\} \varepsilon^*_{\beta}(p-q,s_2)\notag\\
	&=& \bar{u}(q,s_1) \gamma_\mu (\slashed{q}+\slashed{k} + m_{i1}) u_{\nu}(p,s_0)\tilde{g}^{\nu \alpha}_{i2} \tilde{g}^{\mu \beta}_{i3}\Big\{ g_{\beta\lambda}(p-q+k)_\alpha  \notag\\
	&& + g_{\alpha\beta}(p-q-2k)_\lambda  + g_{\lambda\alpha}(k-2p+2q)_\beta \Big\} \varepsilon^{*\lambda}(p-q,s_2),
\end{eqnarray}
\end{widetext}

where $\epsilon_{i1} = m_{i1}\Gamma_{i1}$, $\epsilon_{i2} = m_{i2}\Gamma_{i2}$, $\epsilon_{i3} = m_{i3}\Gamma_{i3}$.
And $ \tilde{g}_{\mu \nu}^{i2} \equiv \tilde{g}_{\mu \nu}(p-q-k,m_{i2}) $, $ \tilde{g}_{\mu \nu}^{i3} \equiv \tilde{g}_{\mu \nu}(k,m_{i3}) $, $ \tilde{g}_{\mu \nu}(p,m) = g_{\mu \nu} - p_\mu p_\nu/m^2 $. 

And $A^{1/2,3/2}_{N^*(2270);K,K^*}$ or $A^{3/2}_{N^*(2080);K,K^*}$ correspond to $p \gamma$ or $p \phi$ channel with $K, K^*$ exchange of triangle diagrams for two-body decays of the $N^*(2270)(1/2^-)$, $N^*(2270)(3/2^-)$ or $N^*(2080)(3/2^-)$.

So, we can calculate the squared amplitude and width by:
\begin{eqnarray}
	|\mM|^2 &=& |\mM_{K}|^2 + |\mM_{K^*}|^2 ,\\[4pt]
    d \Gamma &=& \frac{1}{32 \pi^2} \frac{|\mathcal{M}|^2}{2J+1} \frac{\left|\vec{q}\right|}{m_0^2}  d \Omega .
\end{eqnarray} 

Then we can calculate the corresponding results in Table~\ref{Tab:moleculeresult}.

\begin{table*}[htpb]
	\renewcommand\arraystretch{2.0}
    \centering
    \caption{\label{Tab:moleculeresult}The values of $(|A_{3/2}/A_{1/2}|,\Gamma_{\gamma p},\Gamma_{p\phi})$ at different cutoffs from the triangle diagram.}
    \begin{tabular}{|p{0.35\textwidth}<{\centering}p{0.15\textwidth}<{\centering}|p{0.1\textwidth}<{\centering}|p{0.1\textwidth}<{\centering}|p{0.1\textwidth}<{\centering}|p{0.1\textwidth}<{\centering}|p{0.1\textwidth}<{\centering}|p{0.1\textwidth}<{\centering}|}
    \hline
    \multicolumn{2}{|c|}{$(\Lambda_0,\Lambda_1)(\text{GeV}$)}   & $(0.6,0.8)$ & $(0.8,1.1)$ & $(1.0,1.4)$ & $(1.2,1.7)$ & $(1.4,2.0)$ \\ 
    \hline
    \multicolumn{1}{|c|}{\multirow{3}{*}{$N^*(2080)(3/2^-)$}} & $|A_{3/2}/A_{1/2}|$ & 1.90 & 2.04 & 2.16 & 2.27 & 2.39 \\ 
    \cline{2-7} 
    \multicolumn{1}{|c|}{}                  & $\Gamma_{\gamma p}(\text{KeV})$ & 0.047 & 0.28 & 0.69 & 1.16 & 1.64 \\ 
    \cline{2-7} 
    \multicolumn{1}{|c|}{}                  & $\Gamma_{p\phi}(\text{MeV})$ & 2.34 & 14.97 & 41.09 & 78.70 & 127.98 \\ 
    \hline
    \multicolumn{1}{|c|}{\multirow{3}{*}{$N^*(2270)(3/2^-)$}} & $|A_{3/2}/A_{1/2}|$ & 1.06 & 1.12 &  1.19 & 1.26 & 1.32\\ 
    \cline{2-7} 
    \multicolumn{1}{|c|}{}                  & $\Gamma_{\gamma p}(\text{KeV})$ & 0.53 & 4.56 & 15.65 &  35.56 & 65.72\\ 
    \cline{2-7} 
    \multicolumn{1}{|c|}{}                  & $\Gamma_{p\phi}(\text{MeV})$ & 2.87 & 21.86 & 88.85 & 256.19 & 605.63 \\ 
    \hline
    \end{tabular}
\end{table*}

\end{appendix}

\bibliographystyle{plain}

\end{document}